\documentclass[10pt]{article}

\usepackage{graphicx}
\usepackage{dcolumn}
\usepackage{bm}
\usepackage{subfigure}
\usepackage{enumitem}
\usepackage{amsmath}
\usepackage{amssymb}
\usepackage{amsfonts}
\usepackage{makecell}
\usepackage[export]{adjustbox}

\begin{document}

\begin{center}

{\bf{\large Status on Lattice Calculations of the Proton Spin Decomposition}}

\vspace{0.6cm}


{\bf  Keh-Fei Liu*}

\end{center}

\begin{center}
{
Department of Physics and Astronomy, University of Kentucky, Lexington, KY 40506
} \\
\vspace{0.5cm}
{*E-mail: liu@g.uky.edu}
\end{center}

\begin{abstract}
    Lattice calculations of the proton spin components is reviewed. The lattice results of the
quark spin from the axial-vector current matrix element at $\sim 0.3- 0.4$ is smaller than those from the constituent quark models. This is largely  due to the fact that the vacuum polarization contribution from the disconnected insertion is negative. Its connection with the anomalous Ward identity is clarified and verified numerically. This resolves the contentious issue in the `proton spin crisis'. The glue spin and angular momentum are found to be large and there is notable contribution from the quark orbital angular momentum. Renormalization, mixing and normalization of the quark and glue angular momenta are discussed. With sufficient precision, they can be compared with more precise experimental measurements when the
electron-ion collider facility is available. 
\end{abstract}

\section{\small{Key Words: quark spin, gluon spin, quantum chromodynamics, lattice QCD}}

\section{Introduction} \label{intro}

Prior to the advent of Quantum Chromodynamics (QCD), the structure and mass of hadrons are delineated by the SU(6) non-relativistic quark model. Much like the other non-relativistic systems, where the total spins of atoms and nuclei are the vector
sums of the spins and orbital angular momenta of their constituents, the spin of the proton is from the sum of the spins of the three valence quarks (uud) in the quark model. Thus, it came as a total surprise when it was found in the deep inelastic scattering (DIS) EMC experiment of polarized muon on polarized proton that the quark spin contributes very little to the proton spin ($\Delta \Sigma ({\rm Q^2 = 10\, GeV^2}) = 0.060 \pm 0.047 \pm.069$) ~\cite{Ashman:1987hv,Ashman:1989ig,Hughes:1983kf}. 
The integral of the EMC result on the singlet structure function $g_1 (x, Q^2)$ over $x$ deviates significantly from the Ellis-Jaffe sum rule~\cite{Ellis:1973kp} which is based on the constituent quark model picture. Since this discovery has upended the conventional wisdom at the time, it has sparked spirited discussions and debates. In view of the theoretical quandary and the lack of consensus in the community, this has been dubbed the `proton spin crisis'. The central issue of contention resides on the interpretation of the experimental result -- whether the experimental result is measuring the quark spin alone or the combination of the quark and glue spins. This originated from the implication of the anomalous Ward identity (AWI) which, for the flavor-singlet case, involves a glue topological density term. As will be explained later in Sec.~\ref{quark_spin}, this issued is resolved by the lattice calculation. 

Since the quark spin contribution does not saturate the proton spin (The latest global analysis of experiments give the quark spin at the $\sim 30\%$ level~\cite{deFlorian:2009vb,Nocera:2014gqa,Ethier:2017zbq}), this raises the question as to where the rest of the proton spin resides.  The obvious candidates include glue spin, quark orbital angular momentum, and glue orbital angular momentum. A number of experimental efforts have been carried out or planned to search for these components. These include
the polarized pp experiment at RHIC to extract glue helicity~\cite{Bunce:2000uv,Djawotho:2013pga,Adare:2014hsq}, the
deeply virtual Compton scattering (DVCS) and deeply virtual meson production (DVMP) experiments at
the JLab~\cite{Dudek:2012vr}, HERMES~\cite{Airapetian:2008aa} and COMPASS~\cite{Akhunzyanov:2018nut} to extract quark orbital angular momentum from the measured GPD. Future experiments at the electron ion collider (EIC)~\cite{Accardi:2012qut,Boer:2011fh} will have larger kinematic coverage to go to smaller $x$ and will improve the measurements of quark spin, glue spin and orbital angular momenta. 

There are reviews of the proton spin the readers may wish to consult~\cite{CHENG_1996,Filippone:2001ux,Bass:2004xa,Aidala:2012mv,Leader:2013jra,Ji:2016djn,Deur:2018roz,Ji:2020ena}.  In the present review, we shall concentrate on the lattice calculations of the proton spin components -- quark spin, glue spin, quark orbital angular momentum, and glue angular momentum.

\section{Decomposition of the Proton Spin and Momentum}  \label{decom_spin}

The decomposition of the proton spin and momentum in terms of the quark and
glue contributions can be defined from the forward matrix elements of the QCD energy-momentum tensor.  There
are, in principle, infinite ways to define the decomposition. A meaningful
decomposition will depend on whether each component in the division can be
measured experimentally and it would be desirable that they can be calculated on the 
lattice with either local or non-local operators. 
 
There are two major formulations of the decomposition. One is the Jaffe-Manohar
decomposition~\cite{Jaffe:1989jz}
\begin{equation}  \label{JM}
J = \frac{1}{2} \Delta \Sigma + L_q^{JM} + \Delta{G} + L_G,
\end{equation}
where $ \frac{1}{2} \Delta \Sigma/\Delta{G}$ is the quark/glue spin contribution, and
$L_q^{JM}/L_G$ is the quark/glue orbital angular momentum (OAM)
contribution. This is derived from the canonical energy-momentum tensor in the
light-cone frame with $A^+ =0$ gauge. 
Thus, 
this is superficially gauge dependent and is also frame dependent. Furthermore, while $\Delta{G}$ can be 
extracted from high energy experiments, it had been thought that it cannot be obtained from a matrix
element based on a local operator. This has posed a challenge for the lattice approach for many years. We will
visit this issue in Sec.~\ref{gluespin}.

Another one is the Ji decomposition~\cite{Ji:1996ek}
\begin{equation}  \label{Ji}
J = J_q + J_G = \frac{1}{2} \Delta \Sigma + L_q^{Ji} + J_G,
\end{equation}
where $ \frac{1}{2} \Delta \Sigma$ is the same quark spin contribution as in
Eq.~(\ref{JM}), $L_q^{Ji}$ is the quark OAM, and $J_G$ is the glue angular
momentum (AM). This is derived from the energy-momentum tensor (EMT) in
the Belinfante form and each term in Eq.~(\ref{Ji}) is gauge invariant and frame-independent and can
be calculated on the lattice with local operators.

The intriguing difference between these two decompositions and their respective
realization in experiments have perplexed the community for quite a number of years. 
The partonic picture of the glue spin $\Delta G$ from the gluon helicity distribution~\cite{Manohar:1990jx} and orbital angular momentum (OAM) are naturally depicted in the light-front formalism with $\Delta G$ extractable from high 
energy pp collision and OAM from generalized parton distributions (GPDs) and the Wigner distribution (generalized transverse momentum distributions distribution (GTMD))~\cite{Ji:2003ak,Belitsky:2003nz,Meissner:2009ww,Lorce:2011dv}. Unfortunately, the light-front coordinates are not accessible to lattice QCD calculation since the latter
is based on Euclidean path-integral formulation. To bridge the gap between the light-front formulation
and the lattice calculation, it is shown~\cite{Ji:2013fga} that the matrix elements of appropriate equal-time local operator, when
boosted to the infinite frame, is the same as  those of the gauge-invariant but non-local operator from the light-cone
gluon helicity distribution~\cite{Manohar:1990jx}. 
The proof was first carried out for the glue spin $\Delta G$~\cite{Ji:2013fga} with the local operator 
$\vec{E} \times \vec{A}_{\rm phys}$, where $\vec{A}_{\rm phys}$ is the gauge-invariant part of the gauge potential $A_{\mu}$ and satisfies the non-Abelian transverse condition 
$\mathcal{D}^i A_{\rm phys}^i \equiv \partial^i A_{\rm phys}^i -ig[A^i, A_{\rm phys}^i] = 0$. This is like the transverse gauge-invariant part of the gauge potential $A_{\perp}$ in QED. 
Similar proofs for the $L_q^{JM}$ and $L_G$ as defined from the generalized transverse momentum
distribution (GTMD) are derived with local operators.~\cite{Zhao:2015kca}. It is further shown that the proof for $\Delta G$ in
the Coulomb gauge~\cite{Ji:2013fga} can be generalized to temporal and axial gauges~\cite{Hatta:2013gta}. 

After the usual continuum extrapolation of the lattice results at large but
finite momenta in the $\overline{\rm MS}$ scheme at $\mu$, a large momentum effective field theory (LaMET)
~\cite{Ji:2013fga,Ji:2014gla,Ji:2014lra}, which takes care of the non-commuting UV and $P_z \rightarrow \infty$ 
limits, is suggested to match them to those measured on the light-front. 

     A comparison to QED is in order. Many years of the experimental study of paraxial light beam on matter has been able
to distinguish the different manifestation of the spin and OAM of the beam from the radiation-pressure force (a measure of OAM) and the torque (a measure of spin) on the probed dipole particle~\cite{Bliokh:2014ara}. The separation of spin and OAM is based on the canonical energy-momentum tensor in the physical Coulomb gauge. As we learned earlier, this separation is frame dependent. Since the light is always in the light-front frame, it is the natural frame to define the spin and OAM of the optical beam. Likewise, boosting the proton to the infinite momentum frame makes the weakly interacting gluon partons 
in the proton at high energy behave like the photons in the light beam The transverse size of the proton, analogous to  the width of the light beam, admits the existence of OAM of the gluons in the fast-moving proton.

\section{Quark Spin and Anomalous Ward Identity}  \label{quark_spin}

\begin{figure}[htbp]    
\centering
\vspace*{-2.5cm}
{\includegraphics[width=0.6\hsize, angle = 270]{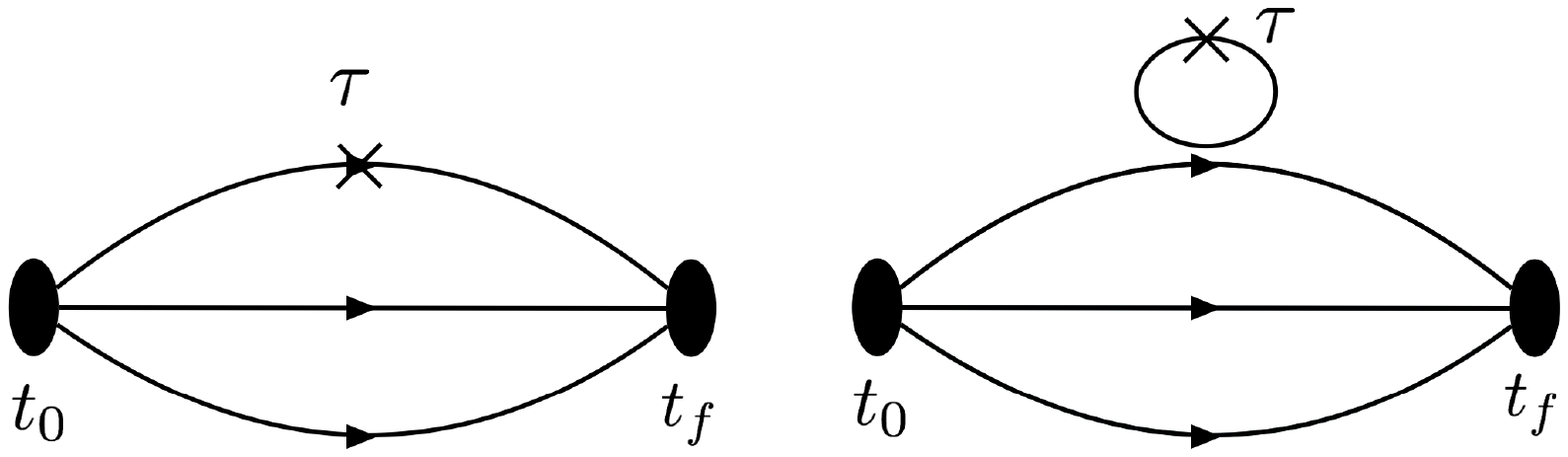}
 }
\vspace*{-2.5cm}
\caption{Three-point function to obtain the nucleon matrix element. Left panel: quark skeleton diagram for the connected insertion (CI) where the current is inserted in the quark propagator between the source at $t_0$ and the sink at $t_f$ for the nucleon. Right panel: disconnected insertion (DI) where the current is inserted to the quark loop which gives the vacuum polarization contribution. The strange quark only contributes in the DI. \label{fig:CIDI}}
 \end{figure}

    Ever since the the emergence of the ``proton spin crisis'', it is incumbent upon lattice QCD, the {\it ab initio} calculation,
 to shoulder the task of understanding the origin of the paltry quark spin contribution. The quark spin contribution can be
 obtained from the nucleon matrix element of the flavor-singlet axial-vector current
 \begin{equation}
  \langle p,s| A_{\mu}^0|p,s \rangle = s_{\mu}\, g_A^0, 
  \end{equation}
 where $A_{\mu}^0 = \sum_{f = u,d,s}\overline{\psi}_f \,i \gamma_{\mu} \gamma_5 \psi_f$ is the flavor-singlet axial-vector current. $s_{\mu}$ is the polarization vector and $g_A^0 = \Delta \Sigma = \Delta u + \Delta d + \Delta s$ is the quark spin contribution
 of the $u,d$ and $s$ quarks. $\langle p,s|p',s\rangle = 2E_p V \delta_{\vec{p},\vec{p'}}$ gives the normalization of the nucleon state.  It involves two parts in the lattice calculation. One is the connected insertion (CI), illustrated
 in the left panel of Fig.~\ref{fig:CIDI} where the current is hooked on the quark line between the source and the sink of
 the nucleon interpolation fields. Another is the disconnected insertion (DI) where the current is coupled to the quark loop
which is a vacuum polarization contribution ({\it N.B.} Only the correlated part between the loop and the nucleon propagator is included. The uncorrelated part is subtracted.)  This is illustrated in the right panel of Fig.~\ref{fig:CIDI}. The strange quark only contributes to the DI. The DI is the most numerically challenging part of the lattice calculation. This is usually calculated
with the noise estimator~\cite{Dong:1993pk}.

 The first lattice calculations with the quenched approximation (without dynamical fermion effects from the fermion 
 determinant)~\cite{Dong:1995rx,Fukugita:1994fh} already revealed that the disconnected insertion (vacuum polarization) in the right panel of Fig.~\ref{fig:CIDI} is negative which reduces that of the CI to make the total smaller than the predicted 
 $g_A^0 = \frac{3}{5} g_A^3 = 0.764$ from the quark model taking the isovector $g_A^3 = 1.2723(23)$ from the muon weak scattering experiment. Over the years, dynamical fermion calculations with physical pion mass, and systematic errors such as continuum and infinite volume limits are beginning to be under control. 
For a recent compilation and evaluation of lattice calculations of $\Delta u, \Delta d$ and $\Delta s$,
one can consult the FLAG Review 2019~\cite{FlavourLatticeAveragingGroup:2019iem}. Here we give a comparison of three recent calculations
(Cyprus group~\cite{Alexandrou:2017oeh}, $\chi$QCD\cite{Liang:2018pis},  and PNDME~\cite{Lin:2018obj}) which have taken into account of multiple lattices at different lattice spacings with light pion masses, some are at the physical one.
They represent the state-of-art calculations on the quark spin contributions for different flavors. They are listed together
with those extracted from experiments. 
\begin{table}[ht]
\centering{}%
\renewcommand{\arraystretch}{1.4}
\begin{tabular}{cccccccc}
 & \textit{$\Delta u$}{ } & \textit{$\Delta d$}{ } & \textit{$\Delta s$}{ } & \textit{$g_A^3$}   & \textit{$\Delta (u+d)$}\!\! (CI)
 & \textit{$\Delta (u/d)$}(DI)& \textit{$\Delta\Sigma$}\tabularnewline
 \hline 
{Cyprus} & {0.830(26)(4)} & {-0.386(16)(6)} & {-0.042(10)(2)} &  {1.216(31)(7)}  & 0.598(24)(6)  &-0.077(15)(5) &{0.402(34)(10)}\tabularnewline
\hline 
{$\chi$QCD} & {0.846(18)(32)} & {-0.410(16)(18)} & {-0.035(8)(7)} & {1.256(16)(30)} & 0.580(16)(30) & -0.072(12)(15)&{0.401(25)(37)}\tabularnewline
\hline
{PNDME} &  {0.777(25)(30)} & {-0.438(18)(30)} & {-0.053(8)} &  {1.218(25)(30)}  & & &{0.286(62)(72)}\tabularnewline
\hline 
{\makecell{de Florian {\it et al.}\\($Q^2$=10 GeV$^2$)}} & {$0.793^{+0.011}_{-0.012}$} & {$-0.416^{+0.011}_{-0.009}$} & {$-0.012^{+0.020}_{-0.024}$} & & & & {$0.366^{+0.015}_{-0.018}$}\tabularnewline
\hline 
{\makecell{NNPDFpol1.1\\($Q^2$=10 GeV$^2$)}} & {0.76(4)} & {-0.41(4)} & {-0.10(8)} & & & & {0.25(10)}\tabularnewline
\hline 
{\makecell{COMPASS\\($Q^2$=3 GeV$^2$)}} & {$[0.82,0.85]$} & {$[-0.45,-0.42]$} & {$[-0.11,-0.08]$} & 1.22(5)(10) & &
&{$[0.26, 0.36]$} \tabularnewline
\hline 
\end{tabular}
\caption{ Results of quark spin for the $u,d$ and $s$ flavors from three recent lattice calculations by the Cyprus group~\cite{Alexandrou:2017oeh},
$\chi$QCD\cite{Liang:2018pis},  PNDME~\cite{Lin:2018obj} in the $\overline{{\rm MS}}$
scheme at 2 GeV are listed. $\Delta (u+d)$(CI) and $\Delta (u+d)$(DI) are the spins of the $u$ and $d$ quarks in the connected insertion\! (CI) and disconnected insertion\! (DI). Three analyses of experiments from de Florian {\it et al.}~\cite{deFlorian:2009vb}, NNPDF~\cite{Nocera:2014gqa} and COMPASS~\cite{Adolph:2015saz} are also listed for comparison.
\label{tab:final_results}}
\end{table}
The total quark spin is composed of contributions from the $u,d$ and $s$ quarks.
\begin{eqnarray}  \label{quark-CD}
\Delta \Sigma  &= & \Delta u +\Delta d + \Delta s , \nonumber \\
   &=& \Delta (u+d) ({\rm CI}) + \Delta (u+d) ({\rm DI}) + \Delta s,
\end{eqnarray}   
where $\Delta (u+d)$\!\! (CI) and $\Delta (u+d)$(DI) are the spins of the $u$ and $d$ quarks in the connected insertion\! (CI) and disconnected insertion \!(DI). $\Delta s$ is the spin of the strange quark. We see that the errors from the lattice calculations are still relatively large, but they are consistent with each other.
They share the same feature in that the DI is negative for all the flavors (i.e., $\Delta (u/d)$\!\! (DI)) for either $u$ or $d$ and
$\Delta s$) which is the reason that the total $g_A^0 \sim 0.3 - 0.4$ is smaller than its CI value of $\sim 0.6$ (see Table~\ref{tab:final_results}).  The lattice calculation of $g_A^0$ requires renormalization and normalization. The non-perturbative renormalization and normalization using the anomalous Ward identity (AWI) are carried out in the calculation of 
$\chi$QCD~\cite{Liang:2018pis}.

As we explained in Sec.~\ref{intro}, to resolve the issue in the `proton spin crisis', one needs to reconcile with the AWI
\begin{equation}  \label{AWI}
\partial_{\mu} A_{\mu}^0 = \sum_{f = u,d,s} 2m_f P_f -   2i N_f q,
 \end{equation}
where $P_f$ is the pseudoscalar density  and $q$ is the topological charge density operator.
The AWI relation is  examined in Ref.~\cite{Liang:2018pis,Yang:2016plb,Gong:2015iir} where the overlap fermion is used.
Overlap fermion is a  chiral fermion that satisfies AWI on the lattice at finite lattice spacing~\cite{Hasenfratz:2002rp}. It is shown numerically that the nucleon matrix elements for Eq.~(\ref{AWI}) is separately satisfied for the CI and DI~\cite{Liang:2018pis}. Therefore, the DI part of $g_A^0$ equals the combined results of the DI of $P_f$ and $q$ on the right-hand-side of Eq.~(\ref{AWI}). In order to calculate the matrix elements in the AWI, one needs to perform a calculation at finite momentum transfer $\vec{q}$ and extrapolate the form factor to $\vec{q} \rightarrow 0$. Such a lattice calculation of the AWI matrix elements has been carried out at $m_{\pi} = 330$ MeV~\cite{Gong:2015iir}. It is found that the forward topological charge matrix element is $\sim 0.15$ for each flavor, which is quite large. The pseudoscalar matrix element is negative with larger magnitude so that the sum of $P_f$ and $q$ in the DI is negative for each flavor. They are equal to $\Delta(u/d)$(DI) and $\Delta s$ as listed in Table~\ref{tab:final_results}.
This proves that the topological charge contribution is part
of the quark spin. This settles the `proton spin crisis' debate as to whether the topological charge contribution is part of
the quark or the glue spin.

\section{Glue Spin}  \label{gluespin}

The recent analyses~\cite{deFlorian:2014yva,Nocera:2014gqa} of the
high-statistics 2009 STAR~\cite{Djawotho:2013pga} and
PHENIX~\cite{Adare:2014hsq} {experiments at RHIC} showed evidence of non-zero
glue helicity in the proton, $\Delta g$. At $Q^2=10$ GeV$^2$, the glue
helicity distribution $\Delta g(x,Q^2)$ is found to be positive and away from
zero in the momentum fraction region $0.05\leq x$ {(specifically $0.05\leq x
  \leq0.2$, the region in which RHIC can determine $\Delta g(x)$ much better
than the other regions)}. However, the results have very large uncertainty in
the region $x\le 0.05$.

Following the suggestion of calculating $\Delta G$ on the lattice as discussed in Sec.~\ref{decom_spin},
a lattice calculation is carried out with the local operator $\vec{S}_G = \int d^3 x\, Tr (\vec{E}\times \vec{A}_{\rm phys})$,
where $\vec{A}_{\rm phys}$ transforms covariantly under the gauge transformation and satisfies the non-Abelian transverse
condition, $\mathcal{D}^i A_{\rm phys}^i = 0$~\cite{Chen:2008ag}. It is shown on the lattice~\cite{Zhao:2015kca} that
$A_{\rm phys}^i$ is related to $A_{c}^i$ in the Coulomb gauge via a gauge transformation, i.e. 
$A_{\rm phys}^{\mu}(x) = g_c (x) A_c^{\mu} g_c^{-1}(x) + \mathcal{O}(a)$, where $g_c$ is the gauge transformation which
fixes the Coulomb gauge. As a result, the glue spin operator
\begin{equation}
\vec{S}_G = \int d^3 x\, Tr (\vec{E}\times \vec{A}_{\rm phys}) = \int d^3 x\, Tr (\vec{E}_c\times \vec{A}_c).
\end{equation}
can be calculated with both $\vec{E}$ and $\vec{A}$ in the Coulomb gauge.
A lattice calculation with the overlap fermion is carried out on 5 lattices with 4 lattice spacings and several sea quark masses including one corresponding to the physical pion mass. The result, when extrapolated to the infinite moment 
limit, gives $\Delta G = 0.251(47)(16)$~\cite{Yang:2016plb} which suggests that the glue spin contributes about 
half of the proton spin. However, there is a caveat. it is found that the finite piece in the one-loop large momentum effective theory (LaMET) matching coefficient is quite large which indicates a  convergence problem for the perturbative series even
after one re-sums the large logarithms. This renders the LaMET matching at current stage not applicable. In this sense,
the glue helicity calculation on the lattice is not completed. 

Another approach is to calculate the polarized glue distribution function $\Delta g(x)$  through the quasi-PDF
approach~\cite{Fan:2018dxu} and  take the first moment to obtain $\Delta G$. The large gauge noise in
this approach poses a challenge.

\section{Orbital Angular Momentum (OAM)}  \label{OAM}

        The Jaffe-Manohar quark and glue orbital angular momenta ($L_q^{JM}$ and $L_G$) can be defined through the
form factor $F_{14}$ of the Wigner distribution function (generalized transverse momentum distribution (GTMD))~\cite{Lorce:2011kd,Hatta:2011ku,Liu:2015xha}. They can be calculated on the lattice with local operators in the nucleon at large momentum and matched to the infinite momentum frame with LaMET~\cite{Zhao:2015kca} as is the case for the glue spin discussed in Sec.~\ref{gluespin}. They can also be calculated directly from GTMD with non-local operators~\cite{Engelhardt:2017miy,Engelhardt:2020qtg}. The Ji's quark OAM $L_q^{Ji}$ can be calculated with the non-local operator in GTMD~\cite{Engelhardt:2017miy,Engelhardt:2020qtg} and can also be obtained
from the gravitational form factors of the EMT. The latter will be discussed in Sec.~\ref{nucleon-spin}. In the approach of Ref.~\cite{Engelhardt:2017miy,Engelhardt:2020qtg}, the longitudinal OAM $L_3$ for the $u$ and $d$ quarks with number
$n$ is defined on the lattice as
\begin{equation}   
\frac{L_3 }{n} = \frac{1}{a} \epsilon_{ij}
\left. \frac{\frac{\partial }{\partial \Delta_{T,j} }
\left( \Phi (a\vec{e}_{i} ) - \Phi (-a\vec{e}_{i} ) \right) }{
\Phi (a\vec{e}_{i} ) + \Phi (-a\vec{e}_{i} )}
\right|_{\Delta_{T} =0}
\label{ratiodef}
\end{equation}
with the proton matrix element defined as
\begin{equation}
\Phi (z_T) = \langle P+\Delta_{T}/2 , S=\vec{e}_{3} |
\overline{\psi}(-z_T /2) \gamma^+ U[-z_T /2,z_T /2]
\psi(z_T /2) | P-\Delta_{T}/2 , S =\vec{e}_{3} \rangle \ .
\label{medef}
\end{equation}       
where $\vec{e}_{3}/\vec{e}_{i} $ is the unit vector in the longitudinal/transverse direction.
The momentum transfer $\Delta_{T} $ and the spatial separation $z_T $ are in the transverse directions and orthogonal
to each other. $U$ is a Wilson line connecting the quark fields at $z_T/2$ and $ - z_T/2$. If it is a
straight line link, it leads to the Ji's OAM. On the other hand, if one uses a staple-shaped link
$U\equiv U[-z/2,\eta v-z/2,\eta v+z/2,z/2]$ with $v$ in the spatial direction and $\eta$ specifies the length
of the staple, the Jaffe-Manohar OAM is realized at $\eta \rightarrow \infty$. The direction of $v$ is
characterized in a  Lorentz invariant parameter $\hat{\zeta} = \frac{v\cdot P}{\sqrt{|v^2 |} \sqrt{P^2}}$, which
needs to be large to approach the light-cone frame. With $v_T = 0$ and $v\equiv -\vec{e}_{3} $,
 $\hat{\zeta } = P_3 /m$ (where $m$ is the proton mass). This implies that a large proton momentum component $P_3$
 is needed to reach a large $\hat{\zeta}$. The partial derivative with respect to $\Delta_{T,j}$ and the finite difference
 of $\Phi (a\vec{e}_{i} ) - \Phi (-a\vec{e}_{i})$ in Eq.~(\ref{ratiodef})  amounts to defining the longitudinal OAM as
 $\vec{L}_3 = \vec{b}_T \times \vec{k}_T $ of the quarks in the proton,  where $b_T $
is the quark impact parameter and $k_T $ the quark transverse momentum.  The $\Phi$ factor in the denominator in
Eq.~(\ref{ratiodef}) cancels the Collins-Soper soft factor in the numerator and serves to normalize $L_3$. 
 
\begin{figure} [tb]
\centering
\subfigure[]
{\includegraphics[width=0.445\hsize]{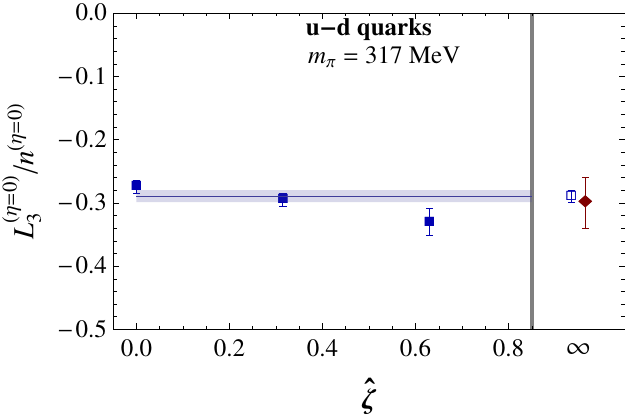}
  \label{Ji_OAM}}
\subfigure[]
{\includegraphics[width=0.445\hsize]{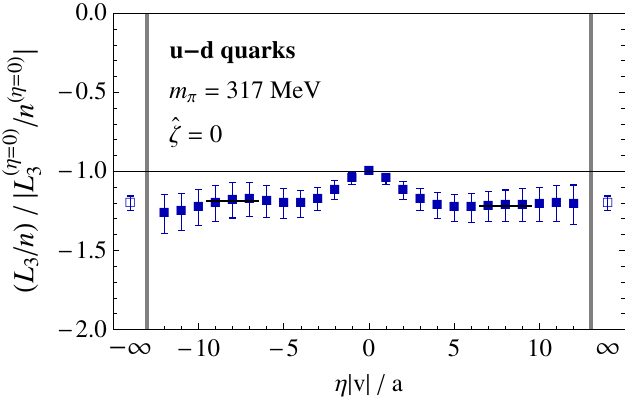}
 \label{JM_eta0} }
\subfigure[]
{\includegraphics[width=0.445\hsize]{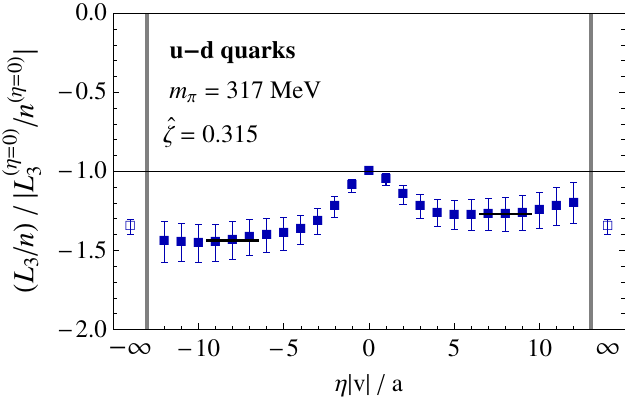}
  \label{JM_eta315}}
 \subfigure[]
{\includegraphics[width=0.445\hsize]{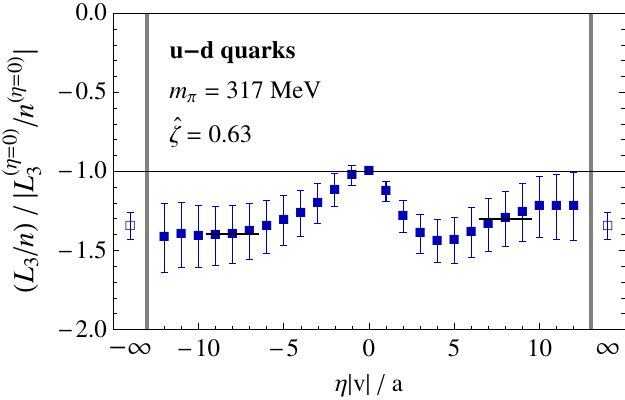}
  \label{JM_eta63} }
\caption{
(a) Ji quark orbital angular momentum, i.e., the $\eta =0$ limit, for the
three values of $\hat{\zeta} $ probed, with the average plotted at
$\hat{\zeta } =\infty $ (open square). The filled diamond represents
the value extracted at the same pion mass in the $\overline{\rm MS} $ scheme at
the scale $\mu^{2} = 4\, \mbox{GeV}^{2} $ via Ji's sum rule in Eq.~(\ref{Ji}).
The isovector $u-d$ quark combination was evaluated. The shown
uncertainties are statistical jackknife errors. (b) Quark orbital angular momentum at $\hat{\zeta}= 0$ as a function 
of staple length parameter $\eta|v|/a $, normalized to the magnitude of Ji quark orbital angular momentum,
i.e., the result obtained at $\eta =0$.  The isovector $u-d$ quark combination is shown with the statistical jackknife errors.
(c) The same as (b) for $\hat{\zeta}=0.315.$ (d) The same as (b) for $\hat{\zeta}=0.63$.}
\label{jiplot}
\end{figure}

     There is a recent lattice calculation on both the Ji OAM and Jaffe-Manohar OAM~\cite{Engelhardt:2020qtg}. 
It is based on the $2+1$-flavor clover fermion ensemble with $m_{\pi} = 317$ MeV. It calculated the Ji OAM for
the isovector $u-d$ quarks from the GTMD in Eq.~(\ref{ratiodef}) with the $U$-link being a straight Wilson line connecting the quark fields in Eq.~(\ref{medef}). It is plotted as a function of $\hat{\zeta}$ in Fig.~\ref{Ji_OAM}. Since Ji OAM is boost-invariant, it is independent of $\hat{\zeta}$.  A constant fit in $\hat{\zeta}$ (open square in Fig.~\ref{Ji_OAM}) shows that it coincides with that obtained from the gravitational form factor (filled diamond in Fig.~\ref{Ji_OAM}) and the sum rule in Eq.~(\ref{Ji}) (to be explained in Sec.~\ref{nucleon-spin}). 
This is a check for consistency between two different approaches. So far, this is a comparison between bare quantities. It would be necessary to have this comparison when renormalization are taken into account in both approaches. 
The Jaffe-Manohar OAM calculations are shown in Figs.~\ref{JM_eta0}, \ref{JM_eta315}, and \ref{JM_eta63} as a  function of 
$\eta |v|/a$  at $\hat{\zeta} = 0, 0.315$ and 0.63. The result from the jointly extrapolated $\eta |v|/a$ and $\hat{\zeta}$ to infinity would correspond to the Jaffe-Manohar OAM. The extrapolation of $\eta |v|/a$ to infinity are shown in Figs.~\ref{JM_eta0}, \ref{JM_eta315}, and \ref{JM_eta63}  by the open squares. The plotted observable is even under $\eta \rightarrow -\eta $, corresponding to time reversal. Accordingly, the $|\eta |\rightarrow \infty $ extrapolated values are obtained by averaging the $\eta >0$ and $\eta <0$ plateaus, which are determined by fitting to the $|\eta | |v|/a =7$ to 9 range. Calculations are carried out for three $\hat{\zeta}$ at  0, 0.315 and 0.63. Since there doesn't seem to be an appreciable variation going
from $\hat{\zeta}$ =0.315 to \mbox{$\hat{\zeta}$ =0.63}, it suggests that the result at $\hat{\zeta}$ =0.315 already well approximates the J-M limit of large $\hat{\zeta}$. It is clear from Fig.~\ref{JM_eta315} that the Jaffe-Manohar OAM at 
$|\eta | |v|/a \rightarrow \infty$ is more negative than that of the Ji OAM at $\eta = 0$. This difference has been interpreted as due to the final state interaction of the quark as it leaves the nucleon target in a DIS experiment~\cite{Burkardt:2012sd}.

This approach can be applied to the glue OAM $L_G$. The present work on the iso-vector quark OAM $L_q^{JM}$ needs to be
renormalized to compare with future experiments in the $\overline{MS}$ scheme. For OAM with different quark flavors, they will mix with $L_G$ and receives mixing contributions from $\Delta \Sigma$ and $\Delta G$~\cite{Ji:2016uqu}. 

It has been shown that $L_q^{JM}$ and $L_G$ can be calculates with local operators as for $\Delta G$ in Sec.~\ref{gluespin}.
However, the matching to the light-cone via LaMET have similarly large matching coefficients~\cite{Ji:2016uqu}.

\section{Quark and Glue Momentum and Angular Momentum Fractions in the Nucleon}  \label{nucleon-spin}

As we see from Sec.~\ref{decom_spin}, besides the quark and glue spins, there are quark and glue orbital angular momenta (OAM) as parts of the proton spin. The OAM can be extracted experimentally from GPD and 
GTMD~\cite{Liu:2015xha}. The calculation of Ji OAM $L_q^{Ji}$ from GTMD, as discussed in Sec.~\ref{OAM},  can 
also be obtained from the gravitational form factors of the energy-momentum tensor (EMT)~\cite{Mathur:1999uf,Hagler:2003jd,Bratt:2010jn}. 

In the following, we shall discuss the calculation of the quark and glue momentum fractions
$\langle x\rangle_q$ and $\langle x\rangle_g$ and the angular momentum decomposition in Eq.~(\ref{Ji}). 
It is shown~\cite{Ji:1996ek} that the momentum and angular momentum can be obtained from the gravitational form factors
of the energy-momentum tensor.
The matrix element of ${\mathcal T}^{\mu\nu}_{q,g}$ between two nucleon states can be 
written in terms of four form factors ($T_1, T_2$, $D$ and  $\bar{C}$). 
\begin{eqnarray} \label{EMT_FF}
\langle P'| (T_{q, g}^{\mu\nu})_R(\mu)|P\rangle /2 M&=& \bar{u}(P')[T_{1_{q,g}}(q^2,\mu) \gamma^{(\mu} \bar{p}^{\nu)} +
T_{2_{q,g}}(q^2,\mu) \frac{\bar{p}^{(\mu} i \sigma^{\nu)\alpha} q_{\alpha}}{2M}   \nonumber \\
     &+&  D_{q,g}(q^2,\mu)\frac{q^{\mu}q^{\nu} - \eta^{\mu\nu}q^2}{M} + \bar{C}_{q,g}(q^2, \mu) M \eta^{\mu\nu} ] u(P),
\end{eqnarray}
where $p$ and $p'$ are the initial and final momenta of the nucleon,\ respectively,\ and 
$\bar{p} = \displaystyle\frac{1}{2}\, (p' + p)$. \mbox{\ $q_\mu = p'_\mu - p_\mu$} is the momentum 
transfer to the nucleon,\ $M$ is the mass of the nucleon. The nucleon spinor \ $u(p,s)$ satisfies the 
normalization conditions $\bar{u}(p,s)\, u(p,s)\, =\, 2m\, , \, $
\mbox{$\displaystyle\sum_s  u(p,s)\, \bar{u}(p,s)\, = \, p\!\!\!/ + m.$}
At the  $q^2 \rightarrow 0$ limit,\ one obtains~\cite{Ji:1996ek}
\begin{eqnarray}    \label{momentum_fraction}
  \label{ang_op_def_split_3}
   \langle x\rangle_{q,g} (\mu) &=& T_1(0, \mu)_{q,g}, \\
   J_{q,g} (\mu)&=& \frac{1}{2} \left[T_1(0, \mu) + T_2(0, \mu)\right]_{q,g}.
\end{eqnarray}
where $\langle x\rangle_{q,g} (\mu)$, obtained from the second moment of the unpolarized PDF, is the momentum 
fraction carried by the quarks or gluons inside a nucleon.\ The other form factor,\ 
$T_2(0, \mu)_{q,g}$,\ is the anomalous gravitomagnetic moment for quarks 
and gluons in an analogy to the anomalous magnetic moment 
$F_2(0)$~\cite{Teryaev:1999su,Brodsky:2000ii}.  Also similar to $F_2(0)$, $T_2(0, \mu)$ cannot be evaluated directly at
$q^2 = 0$ from Eq.~(\ref{EMT_FF}). Instead, it needs to be calculated at small and finite
$q^2$ before taking the $q^2 \rightarrow 0$ limit. This is a much noisier calculation on the lattice than
$T_1(0, \mu)$, particularly for the glue matrix elements. 

By making a connection to the stress tensor of the continuous medium, it is shown~\cite{Polyakov:2002yz,Polyakov:2018zvc} that $D_{q,g}(q^2, \mu)$ is related to the internal force of the hadron and encodes the shear forces and pressure distributions of the quarks and glue in the nucleon. The pressure distribution of the quarks have been deduced from the experimentally measured $D_q(q^2, \mu)$ ~\cite{Burkert:2018bqq} and the pressure distributions for both the quarks and glue from  $D_{q,g}(q^2, \mu)$ are calculated on the lattice~\cite{Shanahan:2018nnv}. The metric term $\bar{C}$ at 
$q^2 = 0$ is equal to the stress term $T^{ii}$ and the sum of the quark and glue parts is zero, i.e.,
$ \bar{C}_q (0)+ \bar{C}_g (0)= 0$ due to the conservation of the EMT, i.e. $\partial_{\nu} T^{\mu \nu}=  0$. Being the
diagonal part of the spatial stress-energy-momentum tensor, $\bar{C}_{q,g}$ has been identified as the pressure~\cite{Lorce:2017xzd,Liu:2021gco}. It is further shown~\cite{Liu:2021gco} from the volume dependence that the gauge part of the trace anomaly gives the negative constant pressure to confine the hadron, the same way Einstein introduced the
cosmological constant in the metric term of the EMT for a static Universe.

 $T_1$ and $T_2$ need to be renormalized and normalized,
similar to the case for the flavor-singlet axial-vector current as discussed in Sec.~\ref{quark_spin}. The perturbative
renormalization of the quarks and glue EMT has been carried out~\cite{Glatzmaier:2014sya,Alexandrou:2016ekb} for 
the quenched~\cite{Deka:2013zha} and $N_f = 2$ calculations~\cite{Alexandrou:2017oeh}. The more reliable and
preferred non-perturbative renormalization with the RI/MOM approach has been done for the $N_f = 2+1$ case~\cite{Yang:2018bft,Yang:2018nqn}, and the $N_f = 2+1+1$ case~\cite{Alexandrou:2020sml}.
The greatest challenge for the non-perturbative renormalization is the renormalization of the glue operator in
the gluon propagator. This involves 3 noisy glue field operators that is as noisy as the glueball calculation. 
Using the Cluster Decomposition Error Reduction (CDER) 
technique~\cite{Liu:2017man}, the glue EMT renormalization is brought under control~\cite{Yang:2018bft}.
CDER takes advantage of the cluster decomposition principle which states that correlator falls off exponentially
with the distance between color-singlet operators. For disconnected insertions, summing over the relative coordinate
between the operators to a limited range when the signal saturates can decrease the error by a $\sqrt{V}$ factor as
compared to summing over the coordinates of both operators, as is usually done. 

Besides renormalization, there are mixing between the quark and glue operators. The structure of the 
renormalization matrix for $\langle x\rangle_q$ and $\langle x\rangle_g$ and for $J_q^{ji}$ and $J_G$ are the same, since they involve the same quark and glue operators.
The mixing between the quark and glue operators are carried out non-perturbatively in the RI/MOM scheme~\cite{Martinelli:1994ty} for the $N_f = 2+1$ flavor calculation with the overlap fermion~\cite{Yang:2018nqn,Yang:2019dha} and perturbatively for the $N_f = 2+1+1$ flavor calculation with the twisted mass
fermion~\cite{Alexandrou:2020sml}.

For the normalization, one relies on the momentum and angular momentum sum rules. In this case,
the normalization conditions for $Z_{T,q}$ and $Z_{T,g}$ are
\begin{eqnarray}   \label{eq:mom_sum_rule}
  \mathcal{N}_q \langle x\rangle_q^{\overline{\rm MS}} (\mu) +  \mathcal{N}_g \langle x\rangle_g^{\overline{\rm MS}} (\mu) &=& 1 , \\
   \mathcal{N}_q J_q^{\overline{\rm MS}} (\mu)+ \mathcal{N}_g J_G^{\overline{\rm MS}} (\mu)&=& \frac{1}{2} . 
  \label{eq:ang_mom_sum_rule}
\end{eqnarray}
It is straight-forward to show~\cite{Ji:1997pf} that Eq.~(\ref{eq:mom_sum_rule}) and Eq.~(\ref{eq:ang_mom_sum_rule}) imply 
\begin{eqnarray}
T_2 (0)_q + T_2 (0)_g &=& 0 .
\label{eq:T_2_sum}
\end{eqnarray}
The vanishing of total $T_2 (0)$, the anomalous gravito-magnetic moment, in the context of a spin-$1/2$ particle was first derived classically from the post-Newtonian manifestation of equivalence principle~\cite{Kobzarev:1962wt}.\ More 
recently,\ this has been proven~\cite{Brodsky:2000ii} for composite systems from the light-front Fock space representation. 

\smallskip

\begin{figure}[htbp]     
\centering
\subfigure[]
{\includegraphics[width=0.42\hsize]{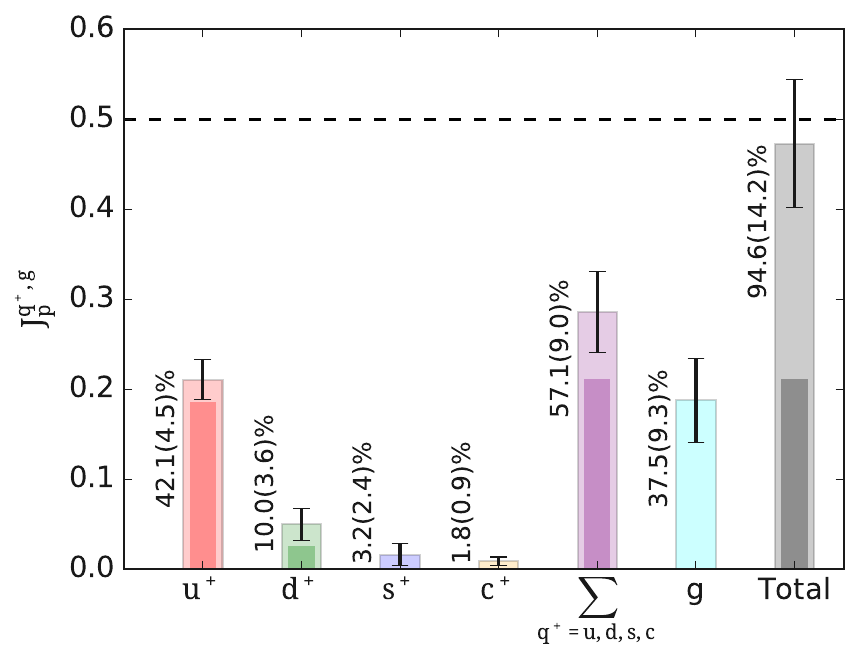}
  \label{ETMC-spin}}
 \hspace{1cm}
\subfigure[]
{\raisebox{40ex}
{\includegraphics[width=0.38\hsize, angle=270]{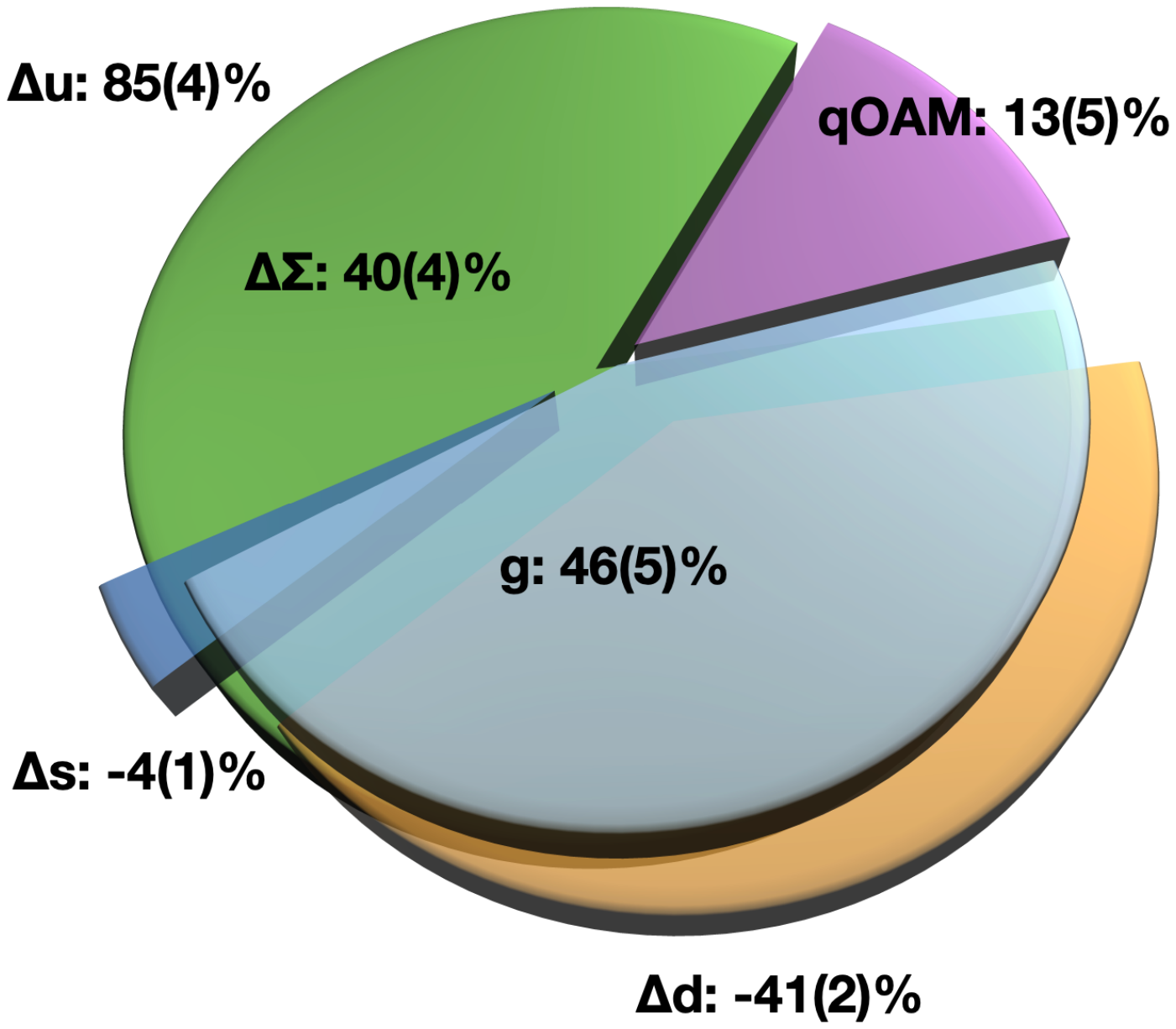}}
  \label{chiQCD-spin}}
\caption{(a) Proton spin decomposition in terms of the angular momentum $J_q$ for the $u,d$ and $s$ quarks and 
the glue angular momentum $J_g$ in Ji's decomposition in the $n_f = 2+1+1$ calculation~\cite{Alexandrou:2020sml}.
 (b) Spin decomposition in terms of the quark spin $\Delta \Sigma$ and its flavor contributions $\Delta u, \Delta d$ and
 $\Delta s$, the glue $J_g$, and the quark OAM for the $n_f = 2+1$ case~\cite{Wang:2021vqy}. }
  \end{figure}
When the normalized and renormalized quark angular momentum $\mathcal{N}_q J_q^{\overline{\rm MS}}$ is calculated, the
quark OAM can be obtained from subtracting the quark spin from it, i.e., $L_q^{Ji} = J_q - \frac{1}{2} \Delta\Sigma$ (cf. Eq.~(\ref{Ji})). The first complete calculation of the quark and glue momenta and angular
momenta in the proton in the gauge invariant formulation (Eq.~\ref{Ji}) was achieved in the quenched approximation with Wilson fermion~\cite{Deka:2013zha,Liu:2011vcs}. The dynamical fermion calculation with $N_f=2$  was done with twisted mass 
fermion~\cite{Alexandrou:2017oeh}. Recently, there are two complete spin decomposition calculations. One is a twisted mass
calculation on a $N_f = 2+1+1$ lattice at $a = 0.08$ fm and pion mass of 139 MeV~\cite{Alexandrou:2020sml}. The quark and glue operators are non-peturbatively renormalized and the quark-glue mixing is carried out perturbatively to one-loop. 
The results on the angular momentum fractions $J$ in Eq.~(\ref{Ji}) for the $u,d,s,c$ quarks and the glue are plotted in Fig.~\ref{ETMC-spin}. Since the total $J$ is consistent with 1/2, no additional normalization is applied. The summed quark $J_q$
is 57.1(9.0)\% and $J_g$ is 37.5(9.3)\% of the total angular momentum. The quark spin contribution is also calculated which
is $\frac{1}{2} \Delta \Sigma = 0.191(15)$. This gives the quark OAM to be 18.8(10.2)(2)\% of the total $J$. Another calculation
is based on the valence overlap fermion on a domain wall fermion sea on a $32^3 \times 64$ lattice at $a = 1.43$ fm and pion mass of 171 MeV with a box size of 4.6 fm~\cite{Wang:2021vqy}. 
The renormalization of the quark and glue operators and their mixings are carried out fully non-perturbatively. 
Since the attempt to calculate $T_2(0)_{q,g}$ did not see a signal beyond one sigma~\cite{Wang:2021vqy},  the normalization is done by assuming the normalization constants for the quark and the glue are the same in the sum rules in Eqs.~(\ref{eq:mom_sum_rule}) and (\ref{eq:ang_mom_sum_rule}). The results~\cite{Wang:2021vqy} on the percentage contributions of 
$\Delta \Sigma$, $J_g$ and $L_q^{ji}$  at 40(4)\%, 46(5)\% and 13(5)\% respectively are shown in Fig.~\ref{chiQCD-spin}. All these results are in the $\overline{\rm MS}$ scheme at 2 GeV. We notice that since the error of the spin components are relatively large in both of these calculations, it is not surprisingly that the results of these calculations are consistent within errors. More work is needed to reduce the statistical errors and control the systematic errors.

\section{Summary}

In the past four decades, lattice calculations have advanced from quenched approximation to including dynamical fermions at the physical quark masses. Multiple lattices with a range of lattice spacings and physical volumes are beginning to be available to allow continuum and infinite volume extrapolations to control systematic errors. 

   The calculations of the quark spin contribution have reached the physical pion point with some
systematic errors taken into account. It is found to contribute $\sim 40\%$ to the proton spin with a $\sim 10\%$ combined statistical and systematic error. It has been demonstrated~\cite{Wang:2021vqy,Alexandrou:2020sml} that all the quark and glue angular momentum components in the Ji decomposition can be tackled in lattice calculations with non-perturbative renormalization and normalization. The glue angular momentum contributes $\sim 38-46\%$ and the quark angular momentum
contribution is in the range $54-57\%$. The errors are at the $\sim 20\%$ level. From the quark spin and the quark angular momentum, one obtains  the quark orbital angular momentum in the range $13 - 18\%$ with an error  of $\sim 50\%$

For the Jaffe-Manohar decomposition, attempts have been made to calculate $\Delta G$~\cite{Yang:2016plb}  with a local operator. However,  there is an unsettled issue in the matching of the glue spin $\Delta G$~\cite{Yang:2016plb}  and orbital angular momentum~\cite{Zhao:2015kca} to the light-front frame. The angular momenta of the quarks and glue can
also be accessed with non-local operators~~\cite{Engelhardt:2017miy,Engelhardt:2020qtg}. In this case, the renormalization
and mixing need to be formulated and carried out. 

With the ongoing experiments at JLab, HERMES and COMPASS on GPD and
TMD and the large kinematic coverage in the future EIC to extend to smaller $x$, the measurements of quark and glue spin
and their angular momentum (total and orbital) are expected to be greatly improved. The lattice calculations of these quantities
with combined statistical and systematic errors less than $\sim 5\%$ are needed to meet the challenge of reliable comparison with experiments in order to gain more insight on the spin structure of the nucleon. 

\section{Acknowledgments}
The author is indebted to M. Engelhardt, X. Ji, C. Lorc\'{e}, Yi-Bo Yang, F. Yuan and Y. Zhao for fruitful discussions. He also thanks Jian Liang, G. Wang and Yi-Bo Yang for the results of the $\chi$QCD Collaboration. This work is partially support by the U.S. DOE grant DE-SC0013065 and DOE Grant No.\ DE-AC05-06OR23177 which is within the framework of the TMD Topical Collaboration. This research used resources of the Oak Ridge Leadership Computing Facility at the Oak Ridge National Laboratory, which is supported by the Office of Science of the U.S. Department of Energy under Contract No.\ DE-AC05-00OR22725. This work used Stampede and Frontera time under the Extreme Science and Engineering Discovery Environment (XSEDE), which is supported by National Science Foundation Grant No. ACI-1053575.
We also thank the National Energy Research Scientific Computing Center (NERSC) for providing HPC resources that have contributed to the research results reported within this paper.
We also acknowledge the facilities of the USQCD Collaboration used for this research in part, which are funded by the Office of Science of the U.S. Department of Energy.

\newpage
\bibliographystyle{unsrt}
\bibliography{Hadron_Structure_Spin}

\end{document}